# EuPdSn$_2$ : magnetic structures in view of resonant x-ray Bragg diffraction

S. W. Lovesey

ISIS Facility, STFC, Didcot, Oxfordshire OX11 0QX, UK

Diamond Light Source, Harwell Science and Innovation Campus, Didcot, Oxfordshire OX11 0DE, UK

Department of Physics, Oxford University, Oxford OX1 3PU, UK

The magnetic properties of materials hosting Eu$^{2+}$ (J = 7/2, 4f$^7$) ions have attracted much attention in the science of strongly correlated electrons. In part because crystal electric field effects are impoverished for an s-state ion, as with Gd$^{3+}$ intermetallics, and Eu$^{2+}$ substitution in biological and optically active materials is resourceful. The magnetic structure of EuPdSn$_2$ is not wholly resolved. Ferromagnetic and antiferromagnetic structures coexist in powder neutron diffraction patterns, and compete in the ground state. Moreover, the specific heat as a function of temperature is enigmatic and indicative of J = 5/2. We present symmetry-informed analytic magnetic structure factors for single crystal resonant x-ray Bragg diffraction using Eu atomic resonances that reveal significant potential for the technique. Europium ions use acentric Wyckoff positions in magnetic space groups inferred from neutron diffraction. In consequence, axial and polar Eu multipoles are compulsory components of both magnetic neutron and resonant x-ray Bragg diffraction patterns. The proposed antiferromagnetic phase of EuPdSn$_2$ supports anapoles (magnetic polar dipoles) already observed in magnetic neutron diffraction patterns presented by Gd doped SmAl$_2$, and several resonant x-ray diffraction patterns.

## 1. Introduction

Interpretations of major studies of the magnetic structure of EuPdSn$_2$ by neutron diffraction on powder samples indicate the presence of both ferromagnetic (FM) and antiferromagnetic (AF) structures (Martinelli *et al*., 2023; Sereni *et al*., 2025). The structures coexist below a temperature ≈ 12 K, and compete in the ground state. The efficacy of neutron diffraction for this compound is curtailed by the high absorption cross-section of natural Eu. Looking ahead, we present symmetry-informed analytic amplitudes for x-ray diffraction by EuPdSn$_2$ with the primary energy tuned to an Eu atomic resonance (Ruck *et al*., 2011; Anderson *et al*., 2017). Bragg diffraction patterns for FM and AF structures are significantly different.

In the theory of resonant x-ray Bragg diffraction used here (Lovesey *et al*., 2005), electronic properties of Eu ions are encapsulated in spherical atomic multipoles of rank K. They represent properties of the magnetic ground-state. Both axial (parity-even) and polar (parity-odd) multipoles are compulsory since Eu ions occupy acentric Wyckoff positions in the FM and AF structures. Non-magnetic monopoles (K = 0) present Thomson scattering at space-group allowed reflections. A magnetic monopole is in possession of the discrete symmetries of a Dirac monopole. The two types of magnetic dipoles (K = 1) are atomic moments, featured in Figs. 1 & 2, and an anapole (Dirac dipole) depicted in Fig. 3 (Scagnoli *et al*., 2011; Lovesey *et al*., 2019). Next in line are quadrupoles (K = 2). A non-magnetic quadrupole creates

Templeton-Templeton scattering from angular anisotropy in the distribution of Eu atomic charge (Templeton & Templeton, 1985). Dirac quadrupoles occur in theories of spintronic and multiferroic materials, e.g., GdCrO$_3$ (Manuel *et al*., 2025; Hayami, 2025).

## 2. Unit cell structure factor

A universal spherical structure factor of rank K,

$$\Psi^K_Q = [\exp(i\mathbf{\kappa} \cdot \mathbf{d}) \langle O^K_Q \rangle_\mathbf{d}], \qquad (1)$$

determines the chemical (nuclear) and magnetic Bragg diffraction patterns for a reflection vector $\mathbf{\kappa}$ defined by integer Miller indices ($h$, $k$, $l$) (Scagnoli & Lovesey, 2009). The implied sum is over Eu ions in sites $\mathbf{d}$. The generic electronic multipole $\langle O^K_Q \rangle$ possesses (2K + 1) projections in the interval $-K \leq Q \leq K$, and its complex conjugate is defined by $(-1)^Q \langle O^K_{-Q} \rangle = \langle O^K_Q \rangle^*$. Angular brackets denote a time-average, or expectation value, of the enclosed spherical tensor operator. Our phase convention for real and imaginary parts labelled by single and double primes is $\langle O^K_Q \rangle = [\langle O^K_Q \rangle' + i \langle O^K_Q \rangle'']$. Cartesian dipole moments in a unit cell ($\xi$, $\eta$, $\zeta$) are $\langle O^1_\xi \rangle = -\sqrt{2} \langle O^1_{+1} \rangle'$, $\langle O^1_\eta \rangle = -\sqrt{2} \langle O^1_{+1} \rangle''$, and $\langle O^1_\zeta \rangle = \langle O^1_0 \rangle$.

Multipoles in resonant x-ray scattering are parity-even ($\sigma_\pi = +1$) for electric dipole-electric dipole (E1-E1, K = 0–2) and electric quadrupole-electric quadrupole (E2-E2, K = 0–4) absorption events. They are time-even $\sigma_\theta = +1$ (time-odd $\sigma_\theta = -1$) for even (odd) rank K, i.e., $\sigma_\theta (-1)^K = +1$. Parity-odd ($\sigma_\pi = -1$) multipoles are compulsory for acentric Wyckoff positions, and $\sigma_\theta = \pm 1$. Multipoles accessed by the parity-odd absorption event E1-E2 have ranks K = 1–3. The structure factor $\Psi^K_Q$ is informed of all elements of symmetry in the magnetic space group. In more detail, equation (1) possesses information about the relevant Wyckoff positions available in the Bilbao table MWYCKPOS for the magnetic symmetry of interest (Bilbao). Site symmetry that might constrain projections Q is given in the same table. Wyckoff positions in a unit cell are related by operations listed in the table MGENPOS (Bilbao). Taken together, the two tables provide all information required to evaluate equation (1) and, thereafter, Bragg diffraction patterns.

The wavelength $\lambda$ of photons with energy E (keV) is $\lambda \approx (12.4/E)$ Å, to a good approximation. Atomic Eu absorption events of immediate interest in resonant x-ray Bragg diffraction by EuPdSn$_2$ include the K edge $\approx$ 48.49 keV, L$_2 \approx$ 7.62 keV, L$_3 \approx$ 6.98 keV (2p $\to$ 5d), and M$_{4,5} \approx$ 1.13 keV (3d $\to$ 4f) (Thole *et al*., 1985; Ruck *et al*., 2011). Cell dimensions for EuPdSn$_2$ are; $a \approx$ 4.4480 Å, $b \approx$ 11.5420 Å, $c \approx$ 7.4266 Å (Martinelli *et al*., 2023). We go on to study reflections ($h$, 0, $l$) and (0, $k$, $l$) for the FM and AF magnetic phases, respectively, and,

$$\text{FM; } \sin(\theta) = (\lambda/2a) [h^2 + (al/c)^2]^{1/2}, \text{ AF; } \sin(\theta) = (\lambda/2a) [k^2 + (al/2c)^2]^{1/2}, \qquad (2)$$

where $\theta$ is the Bragg angle. Factors $(\lambda/2a) \approx 1.23$ and $(\lambda/2c) \approx 0.74$ for M$_{4,5}$, and there are no FM Bragg spots. For the Eu L$_2$ edge $(\lambda/2a) \approx 0.18$ and reflections with even $h$, $l$ are allowed in the FM diffraction pattern. The L$_2$ and L$_3$ edges access 5d and 4f orbitals with E1 and E2 transitions, respectively.

In keeping with standard notation, photon polarizations parallel and perpendicular to the plane of scattering are labelled by π and σ, respectively (Lovesey et al., 2005; Scagnoli & Lovesey, 2009; Paolasini, 2014). Diffraction amplitudes labelled (σ′σ) and (π′π) denote scattering with no rotation of the polarization, e.g., σ → σ′. The two remaining amplitudes (π′σ) and (σ′π) entail the rotation of polarization. In the theory of resonant x-ray diffraction adopted here intensity of a Bragg spot = |(π′σ)|$^2$, for example.

## 3. Ferromagnetic phase (FM)

Axial magnetic dipoles allowed in the FM space group Cm′cm′ (BNS No. 63.464) are depicted in Fig. 1, and a dipole is parallel to the b-axis. Europium ions use Wyckoff positions (4c) with y ≈ 0.4339 (Martinelli et al., 2023). The orthorhombic centrosymmetric magnetic crystal class m′mm′ permits ferromagnetism, a nonlinear magnetoelectric effect, and the piezomagnetic effect. The FM phase is observed in a temperature range ≈ 13.4 K - 10 K (Martinelli et al., 2023), and the corresponding structure factor is,

$$\Psi^K_Q(FM) = \langle O^K_Q \rangle [1 + (-1)^{h+k}][\exp(i\varphi) + \sigma_\pi (-1)^l \exp(-i\varphi)], \quad (3)$$

with φ = 2πky and even (h + k) from the centring condition. Wyckoff position symmetry m′2m′ does not contain inversion. Rotation symmetry elements demand $\sigma_\pi \sigma_\theta (-1)^Q = +1$, and $\langle O^K_Q \rangle = (-1)^{K+Q} \langle O^K_{-Q} \rangle = (-1)^K \langle O^K_Q \rangle^*$. Notably, $\langle O^K_0 \rangle$ is permitted for even K.

Parity-even multipoles $\langle T^K_Q \rangle$ possess a time signature $\sigma_\theta = (-1)^K$, and it leads to even (K + Q). Dipoles K = 1 possess Q = ±1 and are confined to the a-b plane. Bulk magnetic signals, such as XMCD (Lovesey et al., 2005; Anderson et al., 2017), are proportional to $\Psi^K_Q(FM)$ with Miller indices h = k = l = 0. Equation (3) evaluated with $\sigma_\pi$ = +1 and $\sigma_\theta$ = −1 yields $\Psi^1_{+1}(FM) = i4 \langle T^1_{+1} \rangle''$, and bulk ferromagnetism parallel to the b-axis. Dirac multipoles $\langle G^K_Q \rangle$ are revealed in the parity-odd E1-E2 absorption event that requires $\sigma_\pi \sigma_\theta$ = +1, which leads to even Q, and a magnetic monopole $\langle G^0_0 \rangle$.

The fact that $\langle T^K_0 \rangle$ with even rank is permitted in the FM phase means that Thomson scattering $\langle T^0_0 \rangle$ contributes to unrotated diffraction amplitudes (σ′σ) and (π′π) for E1-E1 and E2-E2 events (Scagnoli & Lovesey, 2009). This is not so for the rotated amplitude (π′σ), however. For a reflection vector **κ** = (h, 0, l) with even h, l and an E1-E1 event,

$$(\pi'\sigma) = 2\sqrt{2} \cos(\theta) \cos(\psi) [i\langle T^1_b \rangle + (1/2) \sin(2\beta) \{\sqrt{3} \langle T^2_0 \rangle + \sqrt{2} \langle T^2_{+2} \rangle'\}]$$
$$- 2 \sin(\theta) \sin(2\psi) [\sqrt{(3/2)} \cos^2(\beta) \langle T^2_0 \rangle + (\sin^2(\beta) + 1) \langle T^2_{+2} \rangle']. \quad (4)$$

The azimuthal angle ψ measures rotation of the crystal about **κ**, and the orthorhombic b-axis is normal to the plane of scattering for ψ = 0. The angle β in equation (4) is fixed by cos(β) = h/[h$^2$ + (al/c)$^2$]. Note that (π′σ) is proportional to cos(ψ), and the dipole parallel to the crystal b-axis is 90° out of phase with contributions from quadrupoles. For (0, 0, 2n) Templeton-Templeton scattering (1985) $\langle T^2_{+2} \rangle'$ survives alongside $\langle T^1_b \rangle$.

Dirac multipoles $\langle G^K_Q \rangle$ do not exist in the paramagnetic phase. They are characterized by $\sigma_\pi \sigma_\theta$ = +1, and even Q in the FM phase. An anapole (K = 1) depicted in Fig. 3 does not contribute to reflections ($h$, 0, $l$) with even $h$, odd $l$ using an E1-E2 event. Quadrupole contributions to ($\sigma'\sigma$) and ($\pi'\sigma$) are,

$$(\sigma'\sigma) = (2/\sqrt{15}) \cos(\theta) \cos(\psi) \sin(2\beta) [\sqrt{3} \langle G^2_0\rangle - \sqrt{2} \langle G^2_{+2}\rangle'] + \ldots \quad (5)$$

$$(\pi'\sigma) = (1/\sqrt{5}) [\{\cos(\theta) \cos(\psi)\}^2 - \sin^2(\theta)] [(3\cos^2(\beta) - 1) \langle G^2_0\rangle + \sqrt{6} \sin^2(\beta) \langle G^2_{+2}\rangle']$$
$$- (2/\sqrt{15}) \sin(2\theta) \sin(\psi) \sin(2\beta) [\sqrt{3} \langle G^2_0\rangle - \sqrt{2} \langle G^2_{+2}\rangle'] + \ldots \quad (6)$$

Octupoles (K = 3) are omitted here on the grounds of simplicity; they are readily constructed from available universal expressions (Scagnoli & Lovesey, 2009). There is no quadrupole contribution to ($\sigma'\sigma$) for a reflection (0, 0, $2n$), and ($\pi'\sigma$) reduces to an even function of the azimuthal angle.

## 4. Antiferromagnetic phase (AF)

The AF structure $C_c2/c$ (No. 15.90) is depicted in Fig. 2. Europium ions use Wyckoff positions (8i) at (0.5610, 0, 1/8). A basis {(0, −1, 0), (1, 0, 0), (0, 0, 2)} relative to the parent structure defines orthogonal local axes ($\xi$, $\eta$, $\zeta$) for an Eu ion. The monoclinic centrosymmetric structure belongs to the magnetic crystal class 2/m1′ for which any kind of magnetoelectric effect is prohibited. It is a grey group that contains all three inversions $\bar{1}$, 1′, $\bar{1}'$. Ferromagnetism and the piezomagnetic effect are forbidden. The AF magnetic phase is observed in a temperature range ≈ 12.3 K - 4 K (Martinelli *et al*., 2023), and the corresponding structure factor is,

$$\Psi^K_Q(AF) = \langle O^K_Q\rangle [1 + (-1)^{h+k}] [1 + \sigma_\theta (-1)^l] [\exp(i\gamma) + \sigma_\pi \exp(-i\gamma)], \quad (7)$$

with $\gamma = \{\pi(2hx + l/4)\}$ and x ≈ 0.5610. Magnetic properties are visible for odd $l$, which is a forbidden chemical (nuclear) reflection. A null bulk value of $\Psi^K_Q(AF)$ is correct for antiferromagnetic order. Symmetry of the Wyckoff position (8i) does not include inversion, and rotation elements demand $\langle O^K_Q\rangle = \{\sigma_\pi \sigma_\theta (-1)^{K+Q} \langle O^K_{-Q}\rangle\}$.

For an E1-E1 event $\langle T^K_Q\rangle = \langle T^K_Q\rangle^*$, and $\langle T^K_0\rangle$ is permitted for all K. Allowed axial dipoles are $\langle T^1_0\rangle$ and $\langle T^1_\xi\rangle$, i.e., dipoles are parallel to the orthorhombic c-axis and b-axis. The condition even (K + $l$) follows from the E1-E1 time signature $\sigma_\theta (-1)^K = +1$, and forbidden reflections with odd $l$ are purely magnetic. The E1-E1 amplitude ($\sigma'\sigma$) = 0, because it does not include multipoles with odd K (Scagnoli & Lovesey, 2009). The remaining amplitudes are purely imaginary with a common factor [$i4\sqrt{2} \cos(\pi l/4)$] that is omitted in the results,

$$(\pi'\sigma) = -\cos(\theta) [\cos(\beta) \sin(\psi) \langle T^1_0\rangle + \cos(\psi) \langle T^1_\xi\rangle], \quad (8)$$

$$(\pi'\pi) = \sin(2\theta) [\cos(\beta) \cos(\psi) \langle T^1_0\rangle - \sin(\psi) \langle T^1_\xi\rangle], \quad (9)$$

with $\cos(\beta) = \{(\lambda k)/[2a \sin(\theta)]\}$. The azimuthal angle $\psi$ measures rotation of the crystal sample about the reflection vector (0, $k$, $l$), and the monoclinic $\xi$-axis is normal to the plane of scattering for $\psi = 0$.

Unlike an E1-E1 event, the parity-odd E1-E2 amplitude in the unrotated channel of polarization can be different from zero. It reveals the anapole depicted in Fig. 3 parallel to the orthorhombic a-axis $\langle G^1_\eta \rangle$. Reflections $(0, k, l)$ require even $k$ and odd $l$. At the level of the anapole and quadrupoles the amplitude is,

$$(\sigma'\sigma) = -i8\sqrt{(2/15)} \sin(\pi l/4) \cos(\theta) \sin(\beta) [\sin(\psi) \{(3/\sqrt{10}) \langle G^1_\eta \rangle + \langle G^2_{+1} \rangle'\}$$

$$+ \cos(\beta) \cos(\psi) \{\sqrt{(3/2)} \langle G^2_0 \rangle + \langle G^2_{+2} \rangle'\}] + ... \quad (10)$$

Notably, $\sin(\beta) \propto l$ and it is different from zero for all considered reflections.

## 5. Conclusions

In summary, we present exact analytic amplitudes for resonant x-ray Bragg diffraction from EuPdSn$_2$ using an Eu atomic absorption event. A major study of a powder sample of the compound with magnetic neutron diffraction unveiled ferromagnetic (FM) and antiferromagnetic (AF) phases below a temperature ≈ 12 K depicted in Figs. 1 & 2 (Martinelli *et al.*, 2023). Both phases contribute axial and polar magnetic multipoles to our diffraction patterns. They include rotation of the sample about the reflection vector (an azimuthal angle scan).

Axial dipoles represent atomic magnetic moments in Figs. 1 & 2. Bragg spots in the FM phase satisfy reflection conditions for the parent structure. In consequence, Thomson scattering contributes to diffraction amplitudes in which the orientation of the photon polarization is unchanged, namely, $(\sigma'\sigma)$ and $(\pi'\pi)$. It is absent in the amplitude for rotated polarization $(\pi'\sigma)$ equation (4), which features an axial dipole and Templeton-Templeton scattering (Templeton-Templeton, 1985). Equation (4) is correct for x-ray diffraction enhanced by an electric dipole-electric dipole (E1-E1) absorption event. A corresponding result for the electric quadrupole-electric quadrupole (E2-E2) absorption event is available from the electronic structure factor equation (3) and universal expressions for all diffraction amplitudes (Scagnoli & Lovesey, 2009). Dirac quadrupoles and octupoles (polar magnetic multipoles) are revealed in by the parity-odd E1-E2 absorption event. The corresponding amplitudes equations (5) and (6) produce space-group forbidden Bragg spots. Likewise, all Bragg spots in the AF phase. In this phase, E1-E1 amplitudes $(\pi'\sigma)$ and $(\pi'\pi)$ in equations (8) & (9) contain axial dipoles alone. An anapole contributes to the E1-E2 amplitude $(\sigma'\sigma)$ equation (10), whereas for the same reflection condition and an E1-E1 absorption event $(\sigma'\sigma) = 0$.

**Acknowledgements** Dr D. D. Khalyavin oversaw use of magnetic space groups. Figures 1, 2 and Fig. 3 are supplied by Dr A. Martinelli and Dr V. Scagnoli, respectively.

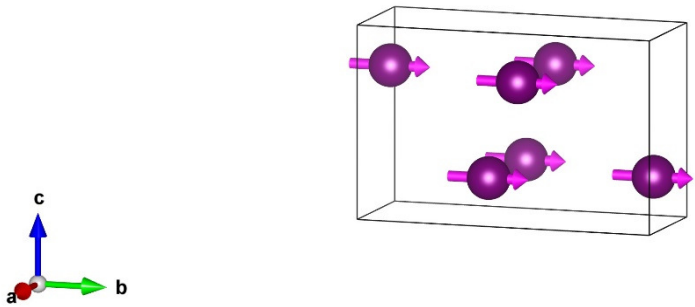

**Figure 1** Ferromagnetic (FM) phase Cm′cm′ (BSN No. 63.464) of EuPdSn$_2$ determined by neutron diffraction and a powder sample. Temperature range ≈ 13.4 K - 10 K (Martinelli *et al.*, 2023).

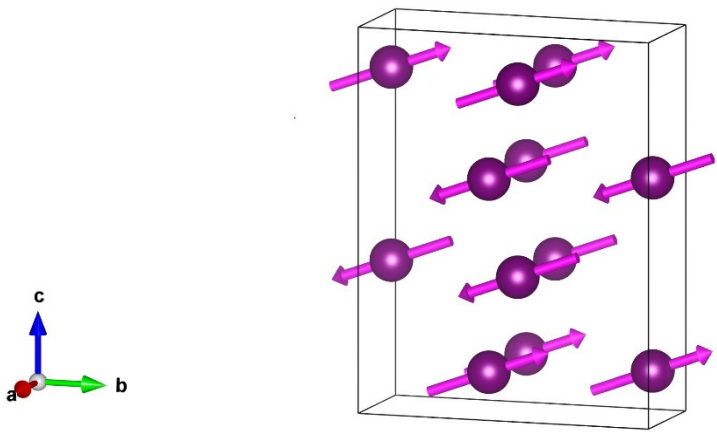

**Figure 2** Antiferromagnetic (AF) phase C$_c$2/c (BSN No. 15.90) of EuPdSn$_2$ in the temperature range ≈ 12. 3 K - 4 K (Martinelli *et al.*, 2023).

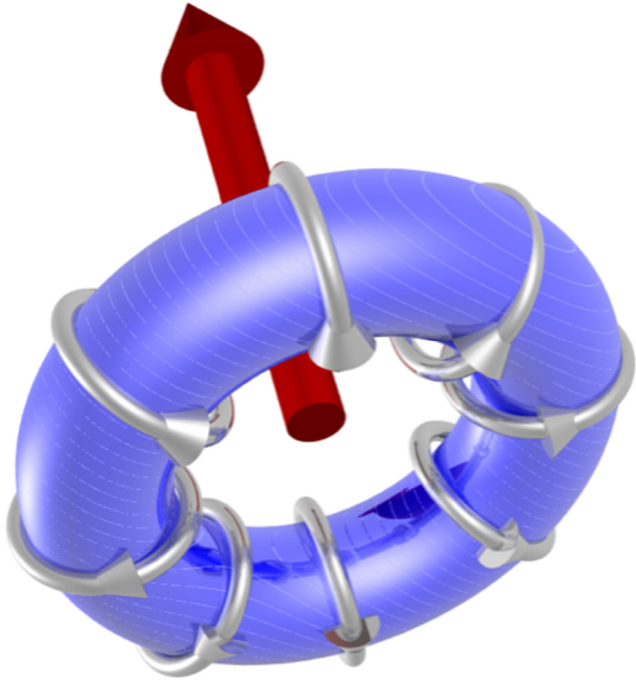

**Figure 3** Depiction of an anapole, also known as a toroidal dipole (Scagnoli *et al.*, 2011).